\documentclass[twocolumn,twoside]{article}

\usepackage{a4}
\usepackage{amssymb}
\usepackage{amsmath}
\usepackage[numbers,sort&compress]{natbib}
\usepackage{graphicx}

\begin{document}

\title{ {\small METHODOLOGICAL NOTES} \\
\vspace{1cm}
{\bf On the anomalous torque applied to a rotating magnetized sphere in a vacuum}}

\date{{\normalsize\textit{P N Lebedev Physical Institute, Russian Academy of Sciences,\\
Leninskii prosp. 53, 119991, Moscow, Russian Federation, \\
Moscow Institute of Physics and Technology (State University),\\
Institutsky per, 9 Dolgoprudny, Russian Federation}}\\[1ex]
{\small \textit{Usp.\ Fiz.\ Nauk} \textbf{184},
865-873 (2014)
[in Russian]\\
English translation: \textit{Physics -- Uspekhi}, \textbf{57}, 799-806 (2014)}
\\{\small Translated by K A Postnov}
}

\author{V.S.Beskin, A.A.Zheltoukhov}

\maketitle

\begin{abstract}
We analyze the torque applied to a rotating magnetized sphere in a vacuum. It is shown that
for the correct determination of one of the torque's component the angular momentum of the
electromagnetic field within the body should be taken into account.
\end{abstract}

\setcounter{secnumdepth}{3}
\setcounter{tocdepth}{2}

\section{Introduction}
As is well known, the first model proposed to describe magnitospheres of pulsars -- rotating 
neutron stars -- was the simplest vacuum model~ \cite{Pacini, OstrikerGunn1969}. According to 
this model, which dates back to the classical paper by Deutsch~\cite{Deutsch1955}, a neutron 
star can be viewed as a highly conducting magnitized solid sphere (with a radius $R$ and a 
magnetic moment $\mathfrak{m}$), rotating in the vacuum with an angular velocity $\Omega$. 
The main power generation occures due to magnetic-dipole{\footnote{The electric quadrupole 
radiation due to charge redistribution within the sphere is supposed by the factor $(\Omega R/c)^{4}$.}} 
radiation, which decelerates the rotation and decreases the angle $\chi$ the spin axis $z^{\prime}$
and magnetic moment $\mathfrak{m}$~\cite{DavisGoldstein1970}. The projection of the breaking  
torque on the spin axis is then expressed as
\begin{equation}
K_{z^{\prime}} = - \frac{2}{3}
\frac{\mathfrak{m}^2}{R^3}\left(\frac{\Omega R}{c}\right)^3 \sin^{2}\chi,
\label{Kmd}
\end{equation}
  and the total power $W_{\rm tot} = - ({\bf \Omega} {\bf K})$ § is~\cite{LL}
\begin{equation}
W_{\rm tot} = \frac{2}{3} \frac{\mathfrak{m}^2 \Omega^{4}}{c^3}\sin^{2}\chi.
\label{Wtot}
\end{equation}
The time evolution of the angle $\chi$ is described by the projection of the torque on the 
$x^{\prime}$ axis lying in the plane $\mathfrak{m}{\bf \Omega}$ which therefore also rotates 
around the $z^{\prime}$ axis with the angular velocity $\Omega$:
\begin{equation}
K_{x^{\prime}} = \frac{2}{3}
\frac{\mathfrak{m}^2}{R^3}\left(\frac{\Omega R}{c}\right)^3 \sin\chi \cos\chi.
\label{Kx}
\end{equation}
It is easy to verify that in this case, the breaking torque ${\bf K}$ is perpendicular 
to the magnetic dipole $\mathfrak{m}$. Therefore, according to Euler's equations, the 
projection of the angular velocity on this axis must be conserved \cite{DavisGoldstein1970}
\begin{equation}
\Omega \cos\chi = {\rm const}.
\label{Ivac}
\end{equation}
As we see, the characteristic time of evolution of the inclination angle  $\chi$ and the 
angular velocity $\Omega$ are the same.

Later, however, it was found that if the neutron star magnitosphere is filled with plasma 
that screens the longitudinal electric field (parallel to the magnetic field), then the 
magnetospheric plasma fully suppresses the magneto-dipole radiation~\cite{BGI, Mestel}.
Energy losses must in this case be related to the action of the Ampere force caused by 
the surface currents that close longitudinal currents in the pulsar magnitosphere; in 
the case of zero longitudinal currents, the total energy loss is zero.

Presently, this statement, which had been aggressively debated for many years after the 
publication of paper~\cite{BGI} in 1983, can be considered to have been proved. For example, 
a numerical solution of the inclined rotator, obtained by Spitkovsky~\cite{Spit} in the 
force-free approximation, does not contain the magneto-dipole wave~\cite{BIP}. We stress 
that, as shown below, the braking of a magnetized sphere rotating in the vacuum is also 
due to the surface currents ~\cite{Michel1991, BGIbook}, but in this case these are purely 
vortex surface currents without sources or sinks.

Here, we do not discuss the model of a magnetosphere filled with plasma, but consider 
the apparently completely studied problem of a rotating magnetized sphere in a vacuum. 
Even in the framework of this simple task, some problems remain open. In particular, 
there is no common opinion regarding the so-called anomalous torque, i.e., the one acting 
along $y^{\prime}$ axis perpendicular to the plane $\mathfrak{m}{\bf \Omega}$ and leading 
not to regular decrease in inclination angle $\chi$ but to the precession of the spin axis.
The name is due to the value of this torque,
\begin{equation}
\label{K}
K_{y^{\prime}} = \xi \frac{\mathfrak{m}^2}{R^3}\left(\frac{\Omega R}{c}\right)^2 \sin\chi \cos\chi,
\end{equation}
where $\xi$ is a numerical coefficient of the order of unity, which turns out to be 
$(\Omega R/c)^{-1}$ times the braking torque $K_{z^{\prime}}$. Here, different authors 
have obtained different values of $\xi$, namely $\xi = 2/3$ ~\cite{OstrikerGunn1969}, 
$\xi = 1$~\cite{DavisGoldstein1970, Goldreich}, $\xi = 1/5$~\cite{GoodNg1985}, and 
$\xi = 3/5$~\cite{Melatos2000} (see also paper~\cite{MestelMoss2005}, in which, however, 
the electric field contribution was admittedly ignored). On the other hand, according 
to~\cite{Michel1991, Istomin}, the anomalous torque is equal to zero ($\xi = 0$), and 
therefore no magnetized sphere precession should occur.

Clearly, such a situation, in which there is no full agreement on the solution of an 
apparently elementary problem, is curios. We recall that the anomalous torque applied 
to a neutron star causes its precession, and for a nonspherical star, this precession, 
superimposed on the deceleration of the radio pulsar rotation, should significantly \
affect the so-called braking index 
$n_{\rm br} = \ddot\Omega \Omega/{\dot \Omega}^2$~\cite{Melatos2000, Barsukov2010, BB}.

Thus, the problem considered is of both theoretical and purely practical interest. 
In this paper, we therefore try to clarify the situation as much we can, and show where 
the variety of results come from. As we see, different papers have in fact discussed 
different quantities, many of which cannot be treated as the torque applied to a magnetized 
sphere rotating in a vacuum. We perform the calculation independently in the framework of 
the so-called quasistationary formalism, which, as we see, allows obtaining the result in 
the fastest and most straightforward way.

\section{Method of calculations}

We first make several general comments. As noted in the Introduction, we are interested 
only in the anomalous torque applied to a magnetized sphere rotating in a vacuum. In 
addiction, an important refinement is in order: as follows from the $z$-component of 
the equation of motion ${\rm d}{\mathfrak{m}}/{\rm d}t = [{\bf \Omega} \times \mathfrak{m}]$
\begin{equation}
\frac{{\rm d}\mathfrak{m}_{z}}{{\rm d}t} = \Omega_{x}\mathfrak{m}_{y} - \Omega_{y}\mathfrak{m}_{x},
\end{equation}
the regular variation of $\mathfrak{m}_{z} = \mathfrak{m} \cos\chi$
related to magnetodipole energy losses is possible if there is a nonzero component of 
the angular velocity ${\bf \Omega}$ lying in the plane $xy$.  However, as follows from 
the same formula, the value of $\Omega_{\perp}$ should be of the order of the inverse 
time of evolution of the angle $\chi$. Therefore, as can be readily verified, this 
additional rotation can be ignored in the analysis of the anomalous torque.

Everywhere below, we assume the solid sphere to be ideally conducting, and hence the 
condition of magnetic field freezing holds everywhere within it:
\begin{equation}
\label{corotat}
{\bf E} + {\bf\beta}_{\rm R}\times{\bf B} = 0,
\end{equation}
where ${\bf \beta}_{\rm R} = {\bf \Omega}\times {\bf r}/c$. 
Clearly, to determine the torque applied to the sphere due to the electromagnetic field, 
it is necessary to calculate the volume and surface currents and changes connected to the 
sphere rotation. As a result, forces applied to the sphere can be presented in the form
\begin{equation}
\label{F-lor}
{\rm d}{\bf F} = \rho_{\rm e} {\bf E}~ {\rm d}V
+ \frac{[{\bf j}\times {\bf B}]}{c}~{\rm d}V
+ \sigma_{\rm e} {\bf E}~ {\rm d}S
+ \frac{[{\bf I}_{\rm S} \times {\bf B}]}{c}~{\rm d}S,
\end{equation}
where the first and second pair of terms in the right-hand side respectively correspond 
to the bulk and surface effects. But if we assume that only corotation currents
\mbox{${\bf j} = c \rho_{\rm e}{\bf \beta}_{\rm R}$} are present within the body (which 
is our key assumption), then it is easy to veryfy that the bulk part of force (\ref{F-lor}) 
vanishes. Now, taking into account that on the sphere ${\bf r} = R {\bf n}$ and 
${\rm d}S = R^2 {\rm d}o$, where ${\rm d}o$ is the solid angle element, we obtain the total 
torque ${\bf K} =\int{\bf r}\times{\rm d}{\bf F}$ in the form
\begin{equation}
\label{Ky}
{\bf K} =
\frac{R^3}{4\pi}\int\Bigl( \left[{\bf n}\times\left\{{\bf B}\right\}\right]
\left({\bf B}{\bf n}\right)+\left[{\bf n}\times{\bf E}\right]
\left(\left\{{\bf E}\right\}{\bf n}\right)\Bigr)
{\rm d}o,
\end{equation}
where the curly brackets denote field jumps on the sphere{\footnote{Here it is important 
that the field components outside the curly brackets are continuous on the sphere.}}. Thus 
the calculation of the torque is reduced to determining the electromagnetic field inside 
and outside the sphere.

We next note that due to the linearity of Maxwell equations, all electromagnetic fields 
can be decomposed into axially symmetric components (with the magnetic axis parallel to 
the spin axis) and orthogonal components. The general solution has the form
\begin{equation}
\label{intro}
A = A^{\parallel}\cos\chi + A^{\perp}\sin\chi,
\end{equation}
where $A$ is an arbitrary field component. Equation (\ref{K}) suggests that only cross 
terms in which one of the component in the products $\left[{\bf n}\times\left\{{\bf B}\right\}\right]
B_{r}š$ and $\left[{\bf n}\times{\bf E}\right] \{E_{r}\}$ 
relates to the orthogonal component make a nonzero contribution to integral (\ref{Ky}).
For example, for a point-like magnetic dipole, the axially symmetric component coincides 
with the static magnetic field
\begin{equation}
\label{Bdip1}
{\bf B}^{\parallel} = \frac{3({\bf \mathfrak{m} n}){\bf n} - \mathfrak{m}}{r^3}.
\end{equation}
For an orthogonal rotator (and again in the case of a point-like dipole) these fields 
must have the form~\cite{Deutsch1955, LL}
\begin{eqnarray}
{B}_{r}^{\perp} & = & \frac{\mathfrak{m}}{r^3} \sin\theta \, {\rm Re} \,
\left(2-2 i \frac{\Omega r}{c}\right) 
\nonumber \\
&&\times \exp\left(i \frac{\Omega r}{c} + i \varphi - i \Omega t \right), 
\label{ll1a} \\
{B}_{\theta}^{\perp} & = & \frac{\mathfrak{m}}{r^3} \cos\theta \, {\rm Re} \,
\left(-1+ i \frac{\Omega r}{c}+\frac{\Omega^2 r^2}{c^2}\right)  \, 
\nonumber \\
&&\times \exp\left( i \frac{\Omega r}{c} +i \varphi - i \Omega t \right),
\label{ll1b}\\
{B}_{\varphi}^{\perp} & = & \frac{\mathfrak{m}}{r^3} \, {\rm Re} \,
\left(-i -\frac{\Omega r}{c}+ i \frac{\Omega^2 r^2}{c^2}\right) \,
\nonumber \\
&&\times\exp\left(i \frac{\Omega r}{c} + i\varphi -i \Omega t \right),
\label{ll1c} 
\end{eqnarray}
\begin{eqnarray}
\quad {E}_{r}^{\perp} & = & 0,\mspace{250mu}
\label{ll2a} \\
{E}_{\theta}^{\perp} & = & \frac{\mathfrak{m}\Omega}{r^2 c} \, {\rm Re} \, 
\left(-1+ i \frac{\Omega r}{c}\right) \,\mspace{100mu} \nonumber\\
&&\times\exp\left( i \frac{\Omega r}{c} + i \varphi -i \Omega t \right),\mspace{40mu}
\label{ll2b} \\
{E}_{\varphi}^{\perp} & = & \frac{\mathfrak{m}\Omega}{r^2 c} \cos\theta\, {\rm Re} \,
\left(- i -\frac{\Omega r}{c}\right) \, \mspace{56mu} \nonumber\\
&&\times\exp\left(i \frac{\Omega r}{c} + i\varphi -i \Omega t \right).\mspace{40mu} 
\label{ll2c}
\end{eqnarray}

Finally, as in most papers, we consider the case of rather slow rotation, where the natural parameter
\begin{equation}
\label{eps}
\varepsilon = \frac{\Omega R}{c}
\end{equation}
is much smaller than unity; for most radio pulsars, $\varepsilon \approx 10^{-3}$--$10^{-4}$.
Comparing expression (\ref{K}) with general relation (\ref{Ky}) for the torque ${\bf K}$,
we then conclude that only the first two terms in the series expansion of the electric and magnetic 
fields in the parameter $\varepsilon$ are needed in order to calculate the anomalous torque.

Therefore, in our opinion, the most convenient method of calculations is the so-called 
quasistationary formalism, which assumes that all fields depend on the azimuthal angle 
$\varphi$ and time $t$ only in the combination $\varphi - \Omega t$. In this case, the 
time derivatives can be substituted by spatial derivatives, and the Maxwell equations 
take the form ~\cite{Book}
\begin{eqnarray}
{\bf\nabla}\times\left({\bf E}
+ {\bf\beta}_{\rm R}\times{\bf B}\right) & = & 0,
\\
{\bf\nabla}\times\left({\bf B}
-{\bf\beta}_{\rm R}\times{\bf E}\right) & = &
\frac{4\pi}{c}{\bf j}-4\pi\rho_e{\bf\beta}_{\rm R}.
\end{eqnarray}
Here, a clear advantage of the proposed method is revealed. Indeed, if we assume the 
corotation condition ${\bf j} = c \rho_{\rm e}{\bf \beta}_{\rm R}$
to be valid inside the sphere, then the right-hand side of Eqn (20)
vanishes both outside the sphere, where currents and charges are absent, and inside 
the sphere. As a result, the following relations should hold both inside and outside 
the sphere:
\begin{eqnarray}
{\bf E} + {\bf\beta}_{\rm R}\times{\bf B} & = & -\nabla \psi,
\label{E}
\\
{\bf B} - {\bf\beta}_{\rm R}\times{\bf E} & = & \nabla h,
\label{B}
\end{eqnarray}
where $\psi(r, \theta, \varphi - \Omega t)$ and $h(r, \theta, \varphi - \Omega t)$ 
are scalar functions that can be found from the condition of continuity of the 
corresponding components od the electric and magnetic field and from the conditions 
${\bf\nabla} \, {\bf E} = 0$ and ${\bf\nabla} \, {\bf B} = 0$ outside the sphere.

Moreover, the proposed method allows obtaining the desired result using a simple 
iteration procedure. Indeed, if the magnetic field $B^{(0)}$ is known in the zeroth 
order in the parameter $\varepsilon = \Omega R/c$, then, using Eqn (\ref{E}), we can 
calculate the electric field $E^{(1)}$ in the first order in the parameter $\varepsilon$.
Equation (\ref{B}), in turn, allows finding the magnetic field $B^{(2)}$ in the second 
order. These two steps are sufficient to calculate the anomalous torque, which is 
proportional, as we have seen, to the square of the small parameter $\Omega R/c$.

Thus, the problem is reduced to finding two scalar functions $\psi^{(1)}$ and $h^{(2)}$, 
that fully determine the structure of electromagnetic fields to the required accuracy. 
Below, we  omit indexes $(1)$ and $(2)$ in most cases.

\section{Results}

We first consider the simplest of a homogeneously magnetized solid sphere. This means 
that in the zeroth order in the parameter $\varepsilon$, the magnetic field is uniform 
inside the sphere and coincides with the field of a point-like dipole outside the sphere. 
Then the zeroth-order magnetic field components inside the sphere have the form
\begin{eqnarray}
B^\perp_r & = & \frac{2{\mathfrak{m}}}{R^3} \sin\theta\cos(\varphi-\Omega t), \quad\nonumber\\
B^\perp_\theta & = & \frac{2{\mathfrak{m}}}{R^3} \cos\theta\cos(\varphi-\Omega t), \quad\nonumber\\
B^\perp_\varphi  & = & - \frac{2{\mathfrak{m}}}{R^3} \sin(\varphi-\Omega t),
\end{eqnarray}
\begin{eqnarray}
B^{\parallel}_r = \frac{2{\mathfrak{m}}}{R^3} \cos\theta, \quad
B^{\parallel}_\theta = -\frac{2{\mathfrak{m}}}{R^3} \sin\theta, \quad
B^{\parallel}_\varphi = 0.
\end{eqnarray}
Correspondingly, outside the sphere, we have
\begin{eqnarray}
{B}^{\perp}_r & = & \frac{2{\mathfrak{m}}}{r^3}\sin\theta\cos(\varphi-\Omega t), \quad\nonumber\\
{B}^{\perp}_\theta & = & -\frac{{\mathfrak{m}}}{r^3}\cos\theta\cos(\varphi-\Omega t), \quad\nonumber\\
{B}^{\perp}_\varphi & = &  \frac{{\mathfrak{m}}}{r^3}\sin(\varphi-\Omega t),
\end{eqnarray}
\begin{eqnarray}
{B}_r^{\parallel} = \frac{2{\mathfrak{m}}}{r^3}\cos\theta, \quad
{B^{\parallel}_\theta} = \frac{{\mathfrak{m}}}{r^3}\sin\theta, \quad
{B^{\parallel}_\varphi} = 0.
\end{eqnarray}

We now turn to the first-order terms in the small parameter $\varepsilon$. We first note 
that in this order, the magnetic field is zero. This, unexpected at first glance, follows 
immediately from relations (\ref{ll1a})--(\ref{ll1c}), where the exponential should be 
expanded in the Taylor series. The magnetic field cannot arise due to $\nabla h$ either, 
since in this order it would correspond to a monopole magnetic field.

As regard the electric field, be comparing Eqns (\ref{corotat}) and (\ref{E}),
we obtain that the condition
\begin{equation}
\psi^{(\rm In)} = 0.
\end{equation}
must always be satisfied inside the sphere. As a result, we have
\begin{eqnarray}
E^{\perp(\rm In)}_r & = & \frac{2{\mathfrak{m}}}{R^3}\,\frac{\Omega r}{c}\,
\sin\theta\cos\theta\cos(\varphi-\Omega t), \quad  \\
E^{\perp(\rm In)}_\theta & = & -\frac{2{\mathfrak{m}}}{R^3}\,\frac{\Omega r}{c}\,
\sin^2\theta\cos(\varphi-\Omega t), \quad
E^{\perp(\rm In)}_\varphi = 0,\nonumber
\end{eqnarray}
\begin{eqnarray}
E^{\parallel(\rm In)}_r & = & - \frac{2{\mathfrak{m}}}{R^3}\,\frac{\Omega r}{c}
\, \sin^2\theta, \quad \\
E^{\parallel(\rm In)}_\theta & = & -\frac{2{\mathfrak{m}}}{R^3}\, \frac{\Omega r}{c}
\, \sin\theta \cos\theta, \quad
E^{\parallel(\rm In)}_\varphi = 0. \nonumber
\end{eqnarray}

We note that for the axially symmetric component, the divergence of the electric 
field is nonzero, which corresponds to a nonzero charge density inside the sphere:
\begin{equation}
 \rho_{\rm GJ} = -\frac{{\bf \Omega} {\bf B}}{2 \pi c}.
\label{GJ}
\end{equation}
This $\rho_{\rm GJ}$ is referred to as the Goldreich-Julian charge density ~\cite{GJ}, 
named after the first to obtain this expression for neutron stars. For the orthogonal 
component, owing to the condition ${\bf \Omega} {\bf B} = 0$, the volume charge 
density inside the sphere is zero.

On the other hand, outside the sphere, according to (\ref{E}) with zero potential 
$\psi = 0$, the electric field must have the form
\begin{eqnarray}
\label{EinOrt}
{E}^{\perp\rm (Out)}_r & = & - \frac{{\mathfrak{m}}}{r^3} \, \frac{\Omega r}{c} \,
\sin\theta\cos\theta\cos(\varphi-\Omega t), \quad \\
{E}^{\perp\rm (Out)}_\theta & = & - \frac{2{\mathfrak{m}}}{r^3}\, \frac{\Omega r}{c}
\sin^2\theta\cos(\varphi-\Omega t), \quad \nonumber \\
{E}^{\perp\rm (Out)}_\varphi & = & 0, \nonumber
\end{eqnarray}
\begin{eqnarray}
\label{EinPar}
{E}_r^{\parallel\rm (Out)} & = &  \frac{{\mathfrak{m}}}{r^3}\,\frac{\Omega r}{c}
\, \sin^2\theta, \quad \\
{E^{\parallel\rm (In)}_\theta}
& = & - \,\frac{2{\mathfrak{m}}}{r^3} \, \frac{\Omega r}{c}
\sin\theta\cos\theta, \quad
{E^{\parallel\rm (Out)}_\varphi} = 0. \nonumber
\end{eqnarray}
In this case, however, it is easy to verify that the electric field divergence is nonzero. 
Therefore, to obtain the divergence-free electric field outside the sphere, where there are 
no charges or currents by definition, these expressions should be corrected using the functions 
$\psi$ in (\ref{E}). It is straightforward to verify that the condition $\nabla \,{\bf E} = 0$ 
for the total field (as well as condition of the continuity of the tangential electric field 
component at the sphere $r = R$) is satisfied for the functions
\begin{eqnarray}
\label{psiMdip0}
\psi^{\perp}_0 & = & \frac{{\mathfrak{m}}}{r} \,
\frac{\Omega}{c} \sin\theta\cos\theta\cos(\varphi-\Omega t)\nonumber\\
&-& \frac{{\mathfrak{m}}}{r^3} \, \frac{\Omega R^2}{c} \sin\theta\cos\theta\cos(\varphi-\Omega t),
\nonumber\\
\psi^{\parallel}_0 & = & -\frac{{\mathfrak{m}}}{r} \, \frac{\Omega}{c} \,\sin^2\theta
+ \frac{1}{3} \, \frac{{\mathfrak{m}}}{r^3} \, \frac{\Omega R^2}{c} (3\cos^2\theta -1).\mspace{32mu}
\end{eqnarray}
Here and below, of course, we use the fact that singularities are absent at the sphere center and 
at infinity (which is also why we have chosen only increasing solutions inside the sphere and 
solutions decrasing at infinity outside it), and that the total charge of the sphere must be 
zero{\footnote{We stress that, as seen from (\ref{E}), the potential
$\psi^{\parallel}_0$ is not the total electric potential of a static axially symmetric problem.}}.

Thus, the electric field outside the sphere takes the form
\begin{eqnarray}
\label{Eoutfull1}
{E}^{\perp\rm (Out)}_r & = & - \frac{3{\mathfrak{m}}}{r^4} \, \frac{\Omega R^2}{c} \,
\sin\theta\cos\theta\cos(\varphi-\Omega t),
\nonumber \\
{E}^{\perp\rm (Out)}_{\theta} & = & - \frac{{\mathfrak{m}}}{r^2}\,
\frac{\Omega}{c} \cos(\varphi-\Omega t)\nonumber\\
&+& \frac{{\mathfrak{m}}}{r^4} \, \frac{\Omega R^2}{c} \,
(1 - 2\sin^2\theta)\cos(\varphi-\Omega t),
\nonumber \\
{E}^{\perp\rm (Out)}_{\varphi} & = & \frac{{\mathfrak{m}}}{r^2}\,
\frac{\Omega}{c} \cos\theta\sin(\varphi-\Omega t)\nonumber\\
&-& \frac{{\mathfrak{m}}}{r^4} \, \frac{\Omega R^2}{c} \, \cos\theta\sin(\varphi-\Omega t),
\end{eqnarray}
\begin{eqnarray}
\label{Eoutfull2}
{E}_r^{\parallel\rm (Out)} & = & -\frac{{\mathfrak{m}}}{r^4}\,
\frac{\Omega R^2}{c} \, (3\cos^2\theta -1),
\nonumber \\
{E^{\parallel\rm (Out)}_\theta} & = & - \frac{2{\mathfrak{m}}}{r^4} \, \frac{\Omega R^2}{c}
\sin\theta\cos\theta,
\nonumber \\
{E^{\parallel\rm (Out)}_\varphi}  & = & 0.
\end{eqnarray}
It is easy to verify that the orthogonal component of the electric field outside the sphere 
is the sum of the magnetic dipole radiation field (\ref{ll2a})--(\ref{ll2c}) and the quadrupole 
field of charges induced in the sphere. The longitudinal field contains only the static field 
of the quadrupole; naturally, this component does not generate electromagnetic waves. Finally, 
jumps of the radial electric field component, which determine the surface charges, are expressed as
\begin{eqnarray}
\{E^{\perp}_{r}\}& = &  -5 \, \frac{{\mathfrak{m}}}{R^3}\,\frac{\Omega R}{c}
\,\sin\theta\cos\theta\cos(\varphi-\Omega t),
\nonumber \\
\{E^{\parallel}_{r}\} & = &  \frac{{\mathfrak{m}}}{R^3}\,\frac{\Omega R}{c} \,(3-5\cos^2\theta).
\end{eqnarray}
Here, the total charge of the shell is nonzero, and the opposite-sign charge related to 
Goldreich-Julian charge density (\ref{GJ}) is uniformly distributed within the sphere volume.

We now determine the second-order fields in $\varepsilon$. Relation (\ref{Ky}) suggests that 
only the magnetic field is relevant here. Indeed, only the magnetic field that appears in 
products with zeroth-order magnetic field contributes to the anomalous torque. Meantime, 
the second-order electric field would contribute to only third-order terms in $\varepsilon$. 
However, as we can again see directly from relations (\ref{ll2a})--(\ref{ll2c}), the electric 
field in this order simply vanishes:
\begin{equation}
{\bf E}^{(2)} = 0.
\label{Ezero}
\end{equation}

The second-order magnetic field can be calculated from the first-order electric field using 
Eqn (\ref{B}). Using the same procedure as for electric fields, it is straightforward to 
find the compensating potentials  $h$, thar is needed for the condition
\mbox{${\bf\nabla} \, {\bf B} = 0$} to be satisfied. As a result, inside the sphere, we obtain
\begin{eqnarray}
\label{hUni0}
h^{\perp({\rm In})} & = & -\frac{3}{5} \, \frac{\mathfrak{m}}{R^3} \,
\frac{\Omega^2 r^3 }{c^2} \, \sin\theta\cos(\varphi-\Omega t), \nonumber\\
h^{\parallel({\rm In})}& = & 0.
\end{eqnarray}
Correspondingly, outside the sphere,
\begin{eqnarray}
\label{Bout2Ortex}
h^{\perp\rm (Out)} & = & \frac{{\mathfrak{m}}}{2} \,
\frac{\Omega^2}{c^2} \,\sin\theta\cos(\varphi-\Omega t)  \\
&-& \frac{{\mathfrak{m}}}{r^2 }\,
\frac{\Omega^2 R^2}{c^2} \,\sin\theta\cos2\theta\cos(\varphi-\Omega t),
\nonumber\\
h^{\parallel\rm (Out)} & = & \frac{{\mathfrak{m}}}{2}\,\frac{\Omega^2}{c^2} \,\cos\theta
+\frac{{\mathfrak{m}}}{r^2}\,\frac{\Omega^2 R^2}{c^2} \,\cos\theta \sin^2\theta.
\nonumber
\end{eqnarray}

Therefore, inside the sphere the second-order magnetic field can be written as
\begin{eqnarray}
\label{Bin2Ort}
B^{\perp\rm (In2)}_r & = & \frac{{\mathfrak{m}}}{R^3}\,\frac{\Omega^2 r^2}{c^2} \,
\sin\theta\left(2\sin^2\theta-\frac{9}{5}\right)\cos(\varphi-\Omega t),
\nonumber\\
B^{\perp\rm (In2)}_\theta & = &  \frac{{\mathfrak{m}}}{R^3}\, \frac{\Omega^2 r^2}{c^2}
\, \cos\theta\left(2\sin^2\theta-\frac{3}{5}\right)\cos(\varphi-\Omega t),
\nonumber\\
B^{\perp\rm (In2)}_\varphi & = & \frac{3}{5} \frac{{\mathfrak{m}}}{R^3}\,
\frac{\Omega^2 r^2}{c^2} \,\sin(\varphi-\Omega t),
\end{eqnarray}
\begin{eqnarray}
\label{Bin2Par}
B_r^{\parallel\rm (In2)}& = &  \frac{2{\mathfrak{m}}}{R^3}\,\frac{\Omega^2 r^2}{c^2}
\,\sin^2\theta\cos\theta,
\nonumber\\
B^{\parallel{\rm (In2)}}_\theta & = & - \frac{2{\mathfrak{m}}}{R^3}\,
\frac{\Omega^2 r^2}{c^2} \,\sin^3\theta,
\nonumber\\
B^{\parallel\rm (In2)}_\varphi & = & 0.
\end{eqnarray}
Correspondingly, outside the sphere we obtain
\begin{eqnarray}
\label{Bout2Ort}
B^{\perp {\rm (Out2)}}_r = \frac{{\mathfrak{m}}}{r}\,\frac{\Omega^2}{c^2}
\sin\theta\cos(\varphi-\Omega t)\nonumber\mspace{115mu}\\
+\frac{{\mathfrak{m}}}{r^3}\,\frac{\Omega^2 R^2}{c^2}
\sin\theta\left(4 \sin^2\theta - \frac{13}{5}\right)\cos(\varphi-\Omega t),
\nonumber\\
B^{\perp {\rm (Out2)}}_\theta = \frac{1}{2} \, \frac{{\mathfrak{m}}}{r}\,
\frac{\Omega^2}{c^2} \cos\theta\cos(\varphi-\Omega t)\nonumber\mspace{95mu}\\
+\frac{{\mathfrak{m}}}{r^3}\,\frac{\Omega^2 R^2}{c^2}
\cos\theta\left(-6\sin^2\theta + \frac{4}{5}\right)\cos(\varphi-\Omega t),
\nonumber\\
B^{\perp {\rm (Out2)}}_\varphi = -\frac{1}{2} \frac{{\mathfrak{m}}}{r}
\,\frac{\Omega^2}{c^2} \,\sin(\varphi-\Omega t)\nonumber\mspace{115mu}\\
+\frac{{\mathfrak{m}}}{r^3}\,\frac{\Omega^2 R^2}{c^2}
\left(\sin^2\theta -\frac{4}{5}\right)\sin(\varphi-\Omega t),
\end{eqnarray}

\begin{eqnarray}
\label{Bout2Q}
B^{\parallel {\rm (Out2)}}_r & = & \frac{4}{5} \,
\frac{{\mathfrak{m}}}{r^3}\,\frac{\Omega^2 R^2}{c^2} \,\cos\theta,
\nonumber\\
B^{\parallel {\rm (Out2)}}_\theta & = & \frac{2}{5} \,
\frac{{\mathfrak{m}}}{r^3}\,\frac{\Omega^2 R^2}{c^2}
\,\sin\theta,
\nonumber\\
B^{\parallel {\rm (Out2)}}_\varphi & = & 0.
\end{eqnarray}
It is easy to verify that the first terms in the orthogonal component (\ref{Bout2Ort})
exactly coincide with the fields of a rotating magnetic dipole, which are proportional 
to $r^{-1}$; to show this, it is again necessary to expand the exponential 
the relations (\ref{ll1a})--(\ref{ll1c}). The second terms correspond to radiation fields 
of quadrupole radiation. As we see, the method we use indeed allows exactly reproducing 
the known results through the second order in $\varepsilon$. As regards the parallel 
component (\ref{Bout2Q}), the second-order magnetic field is simply the field of a magnetic 
dipole equal to $(2/5) \, \varepsilon^2$ times the magnetic dipole of the sphere $\mathfrak{m}$. 
This field is generated by the circle corotation current
${\bf j} = \rho_{\rm e} [{\bf \Omega} \times {\bf r}]$.

The above equations, however, do not yet solve the problem {\footnote{This is already 
seen from the fact that for the normal component, the continuity condition on the sphere 
is not satisfied here.}}. The potentials $h^{(2)}$ are determined up to free harmonic 
functions, which are solutions of the Laplace equation,
\begin{eqnarray}
h^{\rm (In)} & = & \sum_{l=0}^\infty \sum_{m=-l}^{l} f^m_l r^{l}Y^m_l(\theta,\varphi),
\\
h^{\rm (Out)} & = & \sum_{l=0}^\infty \sum_{m=-l}^{l} f^m_l r^{-l-1}Y^m_l(\theta,\varphi),
\end{eqnarray}
where $Y^m_l(\theta,\varphi)$ are spherical functions. Naturally, here again, only solutions 
increasing with $r$ are chosen inside the sphere, and decreasing solutions are chosen outside 
it. This case, as we see, is different from that of the first-order electric field because for 
the potential inside the sphere, the condition $\psi = 0$ was chosen.The continuity of the 
tangential component caused additional harmonic fields to be also absent for $r > R$.

As is easy to verify, the potential $h^{(2)}$ can contain only those spherical functions that 
correspond to the angular distribution of the volume charges currents; therefore, for the orthogonal 
component, we finally obtain
\begin{eqnarray}
\label{hOrt-ab}
h^{\perp\rm(In)} & = & a^{\perp\rm(In)} \frac{\mathfrak{m}}{R^3}
\frac{\Omega^2 R^2}{c^2} \, r \, {\hat Y}^1_1(\theta,\varphi)\nonumber\\
&+&  b^{\perp\rm(In)} \frac{\mathfrak{m}}{R^5}
\frac{\Omega^2 R^2}{c^2}\, r^3 \,{\hat Y}^1_3(\theta,\varphi),
\nonumber\\
h^{\perp\rm(Out)} & = & a^{\perp\rm(Out)} \frac{\mathfrak{m}}{r^{2}}
\frac{\Omega^2 R^2}{c^2}\, {\hat Y}^1_1(\theta,\varphi)\nonumber\\
&+&  b^{\perp\rm(Out)} \frac{\mathfrak{m} R^{2}}{r^{4}}
\frac{\Omega^2 R^2}{c^2}\, {\hat Y}^1_3(\theta,\varphi).
\end{eqnarray}
Correspondingly, for the axially symmetric component, we have
\begin{eqnarray}
\label{hPar-ab}
h^{\parallel\rm(In)} & = & a^{\parallel\rm(In)} \frac{\mathfrak{m}}{R^3}
\frac{\Omega^2 R^2}{c^2}\, r\, {\hat Y}^0_1(\theta,\varphi)\nonumber\\
&+&  b^{\parallel\rm(In)} \frac{\mathfrak{m}}{R^5}
\frac{\Omega^2 R^2}{c^2}\, r^3 \,{\hat Y}^0_3(\theta,\varphi),
\nonumber\\
h^{\parallel\rm(Out)}& = & a^{\parallel\rm(Out)} \frac{\mathfrak{m}}{r^{2}}
\frac{\Omega^2 R^2}{c^2}\,{\hat Y}^0_1(\theta,\varphi)\nonumber\\
&+&  b^{\parallel\rm(Out)} \frac{\mathfrak{m} R^{2}}{r^{4}}
\frac{\Omega^2 R^2}{c^2}\,{\hat Y}^0_3(\theta,\varphi).
\end{eqnarray}
For simplicity, we here use the "nonnormalized" spherical functions
\begin{eqnarray}
{\hat Y}^0_1(\theta,\varphi) & = & \cos\theta,
\nonumber \\
{\hat Y}^0_3(\theta,\varphi) & = & 5\cos^3\theta - 3\cos\theta,
\nonumber \\
{\hat Y}^1_1(\theta,\varphi) & = & \sin\theta\cos\varphi,
\nonumber \\
{\hat Y}^1_3(\theta,\varphi) & = & (5\sin^3\theta - 4\sin\theta)\cos\varphi.
\end{eqnarray}

Thus, the problem is reduced to determining eight coefficients ($a$ and $b$ with 
the various indexes), which are to be found from the normal-component continuity 
on the sphere surface. We hence obtain the following relations between the coefficients:
\begin{eqnarray}
\label{abUni}
a^{\perp\rm(In)} & = &  -2a^{\perp\rm(Out)} +\frac{9}{5}, \nonumber \\
b^{\perp\rm(In)} & = & -\frac{4}{3}\,b^{\perp\rm(Out)}+\frac{2}{15},
\nonumber\\
a^{\parallel\rm(In)} & = & -2a^{\parallel\rm(Out)},\nonumber \\
b^{\parallel\rm(In)} & = & -\frac{4}{3}\,b^{\parallel\rm(Out)}+\frac{2}{15}.
\end{eqnarray}
As we see, relations (\ref{abUni}) are insufficient to find all eight unknown 
coefficients, however. Indeed, to the fields considered in this order, which 
arise due to rotation of the sphere, we can add fields that are formally of 
the order $\varepsilon^2$, but are not related to the rotation itself. Such 
fields can arise due to additional surface currents, not caused by the sphere 
rotation, which are $\varepsilon^2$ times the surface currents generating the 
zeroth-order magnetic field.

The additional fields arising due to potentials (\ref{hOrt-ab})--(\ref{hPar-ab}) 
also contribute to the anomalous torque, and we cannot drop them in the full solution. 
Remarkably, however, the anomalous torque itself is independent od the choice of free 
coefficient.

Indeed, four such free coefficients can be taken to be $(a,b)^{\perp\rm(Out)}$ and 
$(a,b)^{\parallel\rm(Out)}$, which describe harmonic fields outside the sphere.
Direct integration of the corresponding components in the general expression (\ref{Ky})
shows that the anomalous torque is indeed independent of $(a,b)^{\perp\rm(In)}$ and
$(a,b)^{\parallel\rm(Out)}$ because of relations (\ref{abUni}).
Just this must be the case, because if their contribution were nonzero, the contribution 
from the zeroth-ordered term  
$\left[{\bf n}\times\left\{{\bf B}^{(0)}\right\}\right]_{y^{\prime}} B^{(0)}_{r}$ would 
be nonzero, which is also related to free fields described by harmonic functions. On the 
other hand, as can be seen from relations (\ref{abUni}), if all $(a,b)^{\perp\rm(Out)}$ 
and $(a,b)^{\parallel\rm(Out)}$ are set equal to zero, some of the coefficients 
$(a,b)^{\perp\rm(In)}$ and $(a,b)^{\parallel\rm(In)}$ become nonzero and would 
therefore also contribute to $K_{y^{\prime}}$.

To uniquely determine the solution, we again assume that the second-order surface 
currents are solely due to rotation of the surface charge  $\sigma_{\rm e}$:
\begin{eqnarray}
\label{Icond1}
I_{\varphi} & = & \sigma_{\rm e}\Omega R\sin\theta, \\
\label{Icond2}
I_{\theta} & = & 0.
\end{eqnarray}
Conditions (\ref{Icond1}) and (\ref{Icond2}) yield additional relations needed to completely 
determine the coefficients:
\begin{eqnarray}
\label{abUniResult}
&& a^{\perp\rm(Out)}=\frac{7}{30},\quad a^{\perp\rm(In)}= \frac{4}{3},\quad \nonumber\\
&& b^{\perp\rm(Out)}=\frac{1}{7},\qquad b^{\perp\rm(In)}= -\frac{2}{35},
\nonumber\\
&& a^{\parallel\rm(Out)} = 0,\qquad a^{\parallel\rm(In)} = 0, \quad \nonumber\\
&& b^{\parallel\rm(Out)} = \frac{1}{7},\qquad b^{\parallel\rm(In)} = -\frac{2}{35}.
\end{eqnarray}

Importantly, the Deutsch solution ~\cite{Deutsch1955} corresponds to a somewhat different 
problem setup. In ~\cite{Deutsch1955} it was assumed that the normal magnetic field component 
on the sphere does not contain corrections of the order $\varepsilon^2$ at all.
As can be easily verified, this solution corresponds to choosing the constants as
\begin{eqnarray}
\label{abUniResultDeutsch1}
&& a^{\perp\rm(Out)}=-\frac{4}{5},\quad b^{\perp\rm(Out)}=\frac{1}{5}, \quad \nonumber\\
&& a^{\parallel\rm(Out)} = \frac{2}{5}, \qquad b^{\parallel\rm(Out)} = 0,
\end{eqnarray}
whence
\begin{eqnarray}
\label{abUniResultDeutsch2}
&& a^{\perp\rm(In)}= \frac{1}{5},\quad \quad b^{\perp\rm(In)}= -\frac{2}{15}, \quad \nonumber\\
&& a^{\parallel\rm(In)} = -\frac{4}{5}, \qquad b^{\parallel\rm(In)} = \frac{2}{15}.
\end{eqnarray}
Thus, in our setting, the Deutsch solution is the one for a rotation dipole with specially 
adjusted additional sources of small dipole and octupole fields such that they compensate 
the the normal component of the magnetic field of the order  $\varepsilon^2$ on the sphere. 
The value of the anomalous torque, as shown above, is independent of this choice.

We now turn to calculating the anomalous torque itself, Eqn (\ref{Ky}), which can be represented 
in the form
\begin{eqnarray}
\label{Ky2}
 K_{y^{\prime}} =
\frac{R^3}{4\pi}\int\Bigl(
[{\bf n}\times\{{\bf B}^{(2)}\}]_{y^{\prime}} B^{(0)}_{r}  \nonumber\\
 + [{\bf n}\times\{{\bf B}^{(0)}\}]_{y^{\prime}} B^{(2)}_{r}
+[{\bf n}\times{\bf E}^{(1)}]_{y^{\prime}} \{E_{r}^{(1)}\} 
\Bigr) {\rm d}o
\end{eqnarray}
Formula (\ref{Ky2}) can be simplified. Indeed, because the second-order surface current
$I_\varphi = \sigma_{\rm e}\Omega R\sin\theta$ is determined only by the surface charge 
associated with the jump of the first-order electric field, we find
\begin{eqnarray}
\label{rel1}
\{B_{\theta}^{(2)}\} & = & \frac{\Omega R}{c} \sin\theta \{E_{r}^{(1)}\},
\nonumber \\
\{B_{\varphi}^{(2)}\} & = & 0.
\end{eqnarray}
Using (\ref{E}), we can also write
\begin{eqnarray}
\label{rel2}
E_{\theta}^{(1)}  = - \frac{\Omega R}{c} \sin\theta B_{r}^{(0)}
\end{eqnarray}
(this component is continuous on the sphere, and therefore we can set $\psi = 0$),
and hence the first and third terms in (\ref{Ky2}), as can readily be verified, 
cancel each other, and as a result we obtain
\begin{equation}
\label{Ky2last}
K_{y^{\prime}} =
\frac{R^3}{4\pi}\int
[{\bf n}\times\{{\bf B}^{(0)}\}]_{y^{\prime}} B^{(2)}_{r}
{\rm d}o.
\end{equation}
This means that in the absence of zeroth-order surface currents, the anomalous torque is zero.

After performing elementary integration, the total anomalous torque is found to be
\begin{equation}
\label{result1}
{K}_{y^{\prime}} =
\frac{1}{3}\,\frac{\mathfrak{m}^2}{R^3}\left(\frac{\Omega R}{c}\right)^2 \sin\chi \cos\chi.
\end{equation}
The contribution from the surface currents is here given by
\begin{equation}
\label{resultB}
{K}_{y^{\prime}}^B = \frac{\mathfrak{m}^2}{R^3}\left(\frac{\Omega R}{c}\right)^2\sin\chi \cos\chi,
\end{equation}
and contribution from the electric field (i.e., the torque due to surface charges) is
\begin{equation}
\label{resultE}
{K}_{y^{\prime}}^E
= -\frac{2}{3}\,\frac{\mathfrak{m}^2}{R^3}\left(\frac{\Omega R}{c}\right)^2 \sin\chi \cos\chi.
\end{equation}

As the second example, we consider a rotating hollow sphere. In other words, we assume that 
charges and currents, including those that generate the zeroth-order magnetic field, are 
localized in a thin spherical shell with $r = R$. It turns out that this problem does not 
require changing the fields that we found for the orthogonal dipole. Indeed, as noted above 
Goldreich-Julian charge density (\ref{GJ}) for a uniform "horizontal" magnetic field inside 
the sphere is zero. Therefore, at
$r < R$ we can again set $\psi^{\perp} = 0$.

As regards the axially symmetric component, in order to ensure the condition $\rho_{\rm e} = 0$
inside the sphere, the potential
\begin{equation}
\label{d11}
\delta \psi^{\parallel} = -\frac{2}{3}\,\frac{\mathfrak{m}}{R^3}\, \frac{\Omega r^2}{c}.
\end{equation}
must be added to the obtained solution. As a result, only an additional radial electric field 
arises inside the sphere:
\begin{equation}
\label{add12}
\delta E_{r}^{\parallel} = \frac{4}{3}\,\frac{\mathfrak{m}}{R^3}\, \frac{\Omega r}{c}.
\end{equation}
whereas the electric field outside the sphere does not change at all. Here, the electric field 
jump on the surface is expressed as
\begin{eqnarray}
\{E^{\parallel}_{r}\} & = &  \frac{5}{3} \, \frac{{\mathfrak{m}}}{R^3}\,\frac{\Omega R}{c}
\,(1-3\cos^2\theta).
\end{eqnarray}
As we see, the full charge of the shell is zero in this case.

As regards the second-order magnetic field, it can easily be verified that the additional 
electric field (\ref{add12}) gives rise to the additional potential
\begin{equation}
\label{d14}
\delta h^{\parallel} =
\frac{4}{15}\,\frac{\mathfrak{m}}{R^3}\, \frac{\Omega^2}{c^2}\, r^3 \cos\theta.
\end{equation}

As a result, the additional magnetic field inside the sphere, including the free fields, takes the form
\begin{eqnarray}
\label{Binadd}
\delta B^{\parallel {\rm (In2)}}_r & = & -\frac{4}{5} \,
\frac{{\mathfrak{m}}}{R^3}\,\frac{\Omega^2 r^2}{c^2} \,\cos\theta
+ a\frac{{\mathfrak{m}}}{R^3}\,\frac{\Omega^2 R^2}{c^2} \,\cos\theta,
\nonumber\\
\delta B^{\parallel {\rm (In2)}}_{\theta} & = & \frac{8}{5} \,
\frac{{\mathfrak{m}}}{R^3}\,\frac{\Omega^2 r^2}{c^2} \,\sin\theta
- a\frac{{\mathfrak{m}}}{R^3}\,\frac{\Omega^2 R^2}{c^2} \,\sin\theta,
\nonumber\\
\delta B^{\parallel {\rm (In2)}}_\varphi & = & 0.
\end{eqnarray}
Correspondingly, outside the sphere we obtain
\begin{eqnarray}
\label{Boutadd}
\delta B^{\parallel {\rm (Out2)}}_r & = & -\left(\frac{4}{5} + 2a^{\prime}\right) \,
\frac{{\mathfrak{m}}}{r^3}\,\frac{\Omega^2 R^2}{c^2} \,\cos\theta,
\nonumber\\
\delta B^{\parallel {\rm (Out2)}}_{\theta} & = & -\left(\frac{2}{5} + a^{\prime}\right)\,
\frac{{\mathfrak{m}}}{r^3}\,\frac{\Omega^2 R^2}{c^2} \,\sin\theta,
\nonumber\\
\delta B^{\parallel {\rm (Out2)}}_\varphi & = & 0.
\end{eqnarray}
The continuity of the magnetic field normal component and the corotation condition yield
\begin{eqnarray}
\label{aaa}
a = -\frac{2}{9},\quad a^{\prime} = \frac{4}{9}.
\end{eqnarray}
Hence, the full anomalous torque finally becomes
\begin{equation}
\label{result1bis}
{K}_{y^{\prime}} =
\frac{31}{45}\,\frac{\mathfrak{m}^2}{R^3}\left(\frac{\Omega R}{c}\right)^2 \sin\chi \cos\chi.
\end{equation}

Using a similar method, we can also solve the problem in which the uniform zeroth-order 
magnetic field in the parameter $\varepsilon$ occupies only the inner spherical volume 
with a radius  $R_{\rm in}$. For intermediate region $R_{\rm in} < r < R$ (where, as for 
$r < R_{\rm in}$, the potential $\psi$ is zero) and for the region outside the sphere, 
we assume that the zeroth-order magnetic field is that of a point-like dipole. For the 
harmonic functions we then need as many as 16 coefficients, because both increasing and 
decreasing solutions can be taken in the region $R_{\rm in} < r < R$. Eventually, we find
\begin{equation}
{K}_{y^{\prime}} =  \left(\frac{8}{15} -\frac{1}{5}\frac{R}{R_{\rm in}}\right)
\frac{\mathfrak{m}^2}{R^3}\left(\frac{\Omega R}{c}\right)^2 \sin\chi \cos\chi.
\label{result3}
\end{equation}
We see that at $R_{\rm in} = R$, we recover the previous value $\xi = 1/3$.

\section{Discussion}

We have shown that the anomalous torque applied to a rotating magnetized sphere in a 
vacuum is not zero in the general case, and its value depends on the structure of the 
internal electromagnetic field. In particular, we should accept the divergence of the 
anomalous torque as $R_{\rm in}$ tends to zero; however, this situation is unphysical 
and cannot be realized.

We first discuss the results obtained in the previous studies. Unfortunately, 
in~\cite{DavisGoldstein1970, Goldreich}, only the final result $\xi = 1$, is 
presented, which coincides, however, with the braking torque (\ref{resultB})
due to surface currents only. It cannot be ruled out that those papers simply 
ignored the contribution from the electric component in Eqn (\ref{Ky2}). 
Next, we note that there is no direct contradiction with the result in~\cite{Istomin},
where the magnetic field inside the sphere was assumed to be that of a point-like 
dipole; then, as follows from Eqn (\ref{Ky2last}), the contribution from the sphere 
surface to the anomalous torque (which is the only quantity determined in~\cite{Istomin}) 
should be zero. In other papers, as we show below, a quite different quantity was 
considered, which does not have the meaning of a anomalous torque.

Indeed, in virtually all papers discussed above, the anomalous torque was calculated as the 
momentum flux ($K_{i} = - \int \varepsilon_{ijk}r_{j}T_{kl}{\rm d}S_{l}$) of the 
electromagnetic stress tensor $T_{kl}$ using the formula
\begin{eqnarray}
{\bf K}^{M} =\frac{1}{4\pi}\int_S \Bigl(
\left[{\bf r}\times{\bf B}\right]
\left({\bf B}\,{\rm d}{\bf S}\right)+\left[{\bf r}\times{\bf E}\right]
\left({\bf E}\,{\rm d}{\bf S}\right)\nonumber\\
-\frac{1}{2}~({\bf E}^2+{\bf B}^2)\left[{\bf r}\times{\rm d}{\bf S}\right]
\Bigr).\mspace{120mu}
\label{KMelgen}
\end{eqnarray}
When integration over the sphere (with ${\bf r} = R\,{\bf n}$ and
${\rm d}{\bf S} = {\bf n} \,R^2 {\rm d}o$,
where, again, ${\rm d}o$ is the solid angle element), we obtain
\begin{equation}
{\bf K}^{M}_{y^{\prime}} =\frac{R^3}{4\pi}\int\Bigl( \left[{\bf n}\times{\bf B}\right]_{y^{\prime}}
\left({\bf B}{\bf n}\right)+\left[{\bf n}\times{\bf E}\right]_{y^{\prime}}
\left({\bf E}{\bf n}\right)\Bigr) {\rm d}o.
\label{KMel}
\end{equation}
Formula (\ref{KMel}) differs from (\ref{Ky}) in that there are no electric or magnetic 
field jumps on the sphere surface. If we substitute the values of the fields outside 
the sphere as found above in (\ref{KMel}), then at $r = R + 0$ we obtain
\begin{equation}
\xi = \frac{3}{5},
\label{trip}
\end{equation}
which is the result in~\cite{Melatos2000} for the Deutsch solution. Here, it is very 
important that the value $\xi = 3/5$ is also independent of the choice of free coefficients.

However, it should be kept in mind that relation (\ref{KMel}), which indeed can be found 
in many textbooks, is provided with important comments in Landau and Lifshits's \emph{Electrodynamics 
of Continuous Media}~\cite{LL2}. This formula can be used only if the considered volume 'does 
not include charged bodies that are field sources'. Therefore, formula (\ref{KMel}) can be used 
only when the flux of the electromagnetic stress tensor within the body is zero. However, the 
rotating spherical body inside which currents and charges are induced does not satisfy this 
condition, as we now show.

Indeed, the flux of the angular momentum vector of the electromagnetic field is related to 
the torque acting on matter by the electromagnetic field angular momentum conservation 
law~\cite{Schwinger1998}
\begin{equation}
\label{AMconserv}
\frac{{\rm d}\,{\bf L}_{\rm field}}{{\rm d}t}
 + \int [{\bf r}\times {\bf F}] {\rm d}V = {\bf K}^{M}.
\end{equation}
Here, ${\bf L}_{\rm field}$ is the angular momentum of the electromagnetic field inside the volume $V$,
\begin{equation}
{\bf L}_{\rm field} = \int \frac{[{\bf r}\times [{\bf E}\times{\bf B}]]}{4\pi c}{\rm d}V
\label{Lfield}
\end{equation}
${\bf K}^{M}$ is the field angular momentum flux through the surface bounding this volume, and
${\bf F} = \rho_{\rm e}{\bf E} + {\bf j} \times {\bf B}/c$ is the Lorentz force.
The last term in (\ref{AMconserv}) plays the role of a source or a sink and is therefore 
responsible for the angular momentum transfer from the electromagnetic field to matter:
\begin{equation}
\frac{{\rm d\,{\bf L}_{\rm mat}}}{{\rm d}t} = \int [{\bf r}\times {\bf F}] {\rm d}V.
\end{equation}
It is this therm that pays the role of torque, and not ${\bf K}^{M}$, as was assumed 
in~\cite{Melatos2000, Barsukov2010}.

Indeed, consider a sphere of a radius $r < R$ concentric with the body of study.
With the explicit expressions for the zeroth-order magnetic field and the first-order 
electric field, it can be readily verified that for a uniformly magnetized spherical 
body, the time-dependent angular momentum component of ${\bf L}_{\rm field}$ in 
(\ref{Lfield}) has the form
\begin{equation}
{\bf L}_{\rm field} = \frac{4}{15}\,\frac{\mathfrak{m}^2}{R^6}
\left(\frac{\Omega r^5}{c^2}\right) \sin\chi \cos\chi \, {\bf e}_{x^{\prime}}.
\label{Lff}
\end{equation}
As the radius $r$ increases, ${\bf L}_{\rm field}$ continuously increases, and hence 
the angular momentum flux ${\bf K}^{M}$ related to rotation of this vector increases; 
it is discontinuous only on the sphere \mbox{$r=R$}. Because 
${\dot {\bf e}_{x^{\prime}}} =\Omega \, {\bf e}_{y^{\prime}}$, the time derivative 
${\dot {\bf L}_{\rm field}} = \Omega L_{\rm field} {\bf e}_{y^{\prime}}$ exactly 
corresponds to the angular momentum flux ${\bf K}^{M}$, calculated for the inner 
surface of the sphere $r = R - 0$ ($4/15 = 3/5 - 1/3$).

Thus we arrive at an important conclusion: the force balance taken into account in 
the second order in $\varepsilon$, the electromagnetic field angular momentum should 
also be taken into account. Therefore, part of the stresses due to the electromagnetic 
field should affect the angular momentum of the field itself, and only the remaining 
part should influence the interaction with the rotating body. According to Eqn(\ref{AMconserv}), 
this implies that for the torque acting on the sphere, we should use the expression
\begin{equation}
{\bf K} \equiv \frac{{\rm d} {\bf L}_{\rm mat}}{{\rm d}t} = \left\{{\bf K}^{M}\right\}.
\end{equation}

Incidentally, the divergence $\sim R_{\rm in}^{-1}$ arising in Eqn(\ref{result3}) now 
becomes clear. Indeed, the direct calculation of the total angular momentum of the 
electromagnetic field inside the body in this case yields
\begin{equation}
{\bf L}_{\rm field} =  \frac{\mathfrak{m}^2 \Omega}{c^2 R}
\left(\frac{1}{15} + \frac{1}{5}\, \frac{R}{R_{\rm in}}\right) \sin\chi \cos\chi \, {\bf e}_{x^{\prime}}.
\label{Lfff}
\end{equation}
As we see, the angular momentum of the field contained inside the sphere of radius $R$
at given value of $\mathfrak{m}$ diverges as $1/(5 R_{\rm in})$. Therefore, the scale 
of forces applied to the rotating spherical body must be of the same magnitude but with 
the opposite sign.

In conclusion, we stress that the method proposed here is inapplicable to the calculation 
of the torque that is responsible for magneto-dipole radiation. As can be seen from (\ref{Kmd}),
this torque must be of the third order in $\varepsilon$. Therefore, to determine this torque, 
the magnetic field $B^{(3)}$ in the third order in $\varepsilon$ must be known 
(the electric field, owing to condition (\ref{Ezero}) does contribute in the third order). 
These fields, involved in products with $B^{(0)}$, must lead to the required value of
${\bf K}$ (\ref{Kmd}).

But as can be easily verified, the third-order magnetic field $B^{(3)}$ is simply a uniform 
field, whose value cannot be determined by our procedure~\cite{BGIbook}{\footnote{Apparently, 
this is because the method we use here does not discriminate between the retarded and advanced 
potentials~\cite{Book}.}}. Fortunately, this uncertainty arises only at the next step of the 
expansion, because, as we have seen, the anomalous torque (\ref{Ky}) is $(\Omega R/c)^{-1}$ 
times the braking torque directed against the spin axis. As a result, the procedure described 
above is applicable to the problem posed.

On the other hand, if we take the uniform third-order magnetic field from explicit expressions 
for a rotating point-like dipole (\ref{ll1a})--(\ref{ll1c})
\begin{equation}
{\bf B}^{(3)} = -\frac{2}{3} \, \frac{\mathfrak{m}}{R^3}\left(\frac{\Omega R}{c}\right)^3
{\bf e}_{y^{\prime}},
\end{equation}
then the direct calculation of the electromagnetic angular momentum flux ${\bf K}^{M}$ at 
$r = R - 0$ yields $K_{x^{\prime}} = 0$ and $K_{z^{\prime}} = 0$. At $r = R + 0$ naturally, 
we return to expressions (\ref{Kmd}) and (\ref{Kx}). This means that in the third order in 
$\varepsilon$, the electromagnetic field angular momentum within the body is zero. Therefore, 
to this order, the torque applied to a rotating sphere can indeed be determined in terms of 
the surface integral that has no field jumps. By contrast, as follows from numerical 
simulations \cite{TPL}, the flux $K_{x^{\prime}}$ at $r > R$ depends on the integration 
radius. This means that outside the body, the third-order electromagnetic field angular 
momentum is also nonzero.

\section{Conclusion}

The anomalous torque $K_{y^{\prime}}$ acting on a rotating magnetized sphere in the general 
case is indeed nonzero. However, $K_{y^{\prime}}$ depends on the internal structure of the 
fields because in the second order in the parameter $\varepsilon$ the electromagnetic field 
angular momentum ${\bf L}_{\rm field}$ in (\ref{Lfield}) must be taken into account in the 
balance of forces, and this angular momentum in turn depends on the internal electric field 
structure. The result obtained in the three examples considered in Section 4 are different 
because, having the same normal magnetic field component $B_{r}^{(0)}$ on the sphere, each 
case has a different internal electromagnetic field structure. This result in different angular 
momenta of the electromagnetic field. But, in the third order in $\varepsilon$, the electromagnetic 
field angular momentum inside the body is zero. Therefore, when calculating the torques 
$K_{x^{\prime}}$ and $K_{z^{\prime}}$, we can use the angular momentum flux
$K^{M}_{i}$ in Eqn (\ref{KMelgen}), which does not have field jumps on the surface.

At last, following Archimedes, we can cry "Eureka!" Indeed, the measurement of the anomalous 
torque applied to a rotating spherical body allows determining its internal structure, which 
has no apparent manifestations in the outer regions. As we have shown here, lower-order electromagnetic 
fields outside a solid and hollow sphere must be the same, while the applied torques are 
different by a factor of more than two. This is not the first such example in electrodynamics, 
however. For instance, if a body has the so-called anapole moment {\footnote{An anapole moment 
can be modeled by a toroidal solenoid with a poloidal winding current $I$.}}
\begin{equation}
{\bf T} = \frac{1}{10 c} \int [({\bf j}{\bf r}) {\bf r} - 2 r^2 {\bf j}] {\rm d}V,
\end{equation}
then, in the absence of rotation, the electromagnetic fields outside the body are exactly equal 
to zero. But in a nonuniform magnetic field, the torque 
\mbox{${\bf K} = [{\bf T} \times [\nabla \times {\bf B}]]$} acts on the body, and the rotation 
of the body is accompanied by electromagnetic radiation~\cite{anapole}. Taking the field angular 
momentum into account is also absolutely necessary in some other cases (see, e.g.
~\cite{sokolov, chrip} and references therein).

The authors thank D.P.Barsukov, Ya.N.Istomin, D.N.Sobyanin, A.A.Filippov, and A.D.Tchekhovskoy 
for the fruitful discussions, and also A.K. Obukhova and E.E. Stroynov for the help in calculations. 
This paper was supported by the RFBR grant No. 14-02-00831.


\begin{thebibliography}{99}
\bibitem{Pacini}
Pacini F {\it Nature} London {\bf 221} 567 (1967)
\bibitem{OstrikerGunn1969}
 Ostriker J P, Gunn J E {\it ApJ} {\bf 458} 347 (1969)
\bibitem{Deutsch1955}
 Deutsch A J {\it Annales d'Astrophysique} {\bf 18} 1 (1955)
\bibitem{DavisGoldstein1970}
 Davis L, Goldstein M {\it ApJ} {\bf 159} L81 (1970)
\bibitem{LL}
Landau L D, Lifshits E M {\it The Clssical Theory of Fields} (Oxford:Pergamon Press, 1973)
\bibitem{BGI}
Beskin V S' Gurevich A V' Istomin Ya N {\it Sov. Phys. JETP} {\bf 58} 235 (1983)
\bibitem{Mestel}
Mestel L, Panagi P, Shibata S \, {\it Mon. Not. Roy. Astron. Soc.} {\bf 309} 388 (1999)
\bibitem{Spit}
Spitkovsky A  {\it ApJ} {\bf 648} L51 (2006)
\bibitem{BIP}
Beskin V S, Istomin Ya N, Philippov A A {\it Phis. Usp.} {\bf 56} 164 (2013)
\bibitem{Michel1991}
Michel F C \emph{Theory of neutron star magnetospheres} (Univ. of Chicago Press, 1991)
\bibitem{BGIbook}
Beskin V S, Gurevich A V, Istomin Ya N {\it Physics of the pulsar magnetosphere}
(Cambridge Univ. Press, 1993)
\bibitem{Goldreich}
Goldreich P {\it ApJ} {\bf 160} L11 (1970)
\bibitem{GoodNg1985}
Good M L, Ng K K {\it ApJ} {\bf 299} 706 (1985)
\bibitem{Melatos2000}
 Melatos A {\it Mon. Not. Roy. Astron. Soc.} {\bf 313} 217 (2000)
\bibitem{MestelMoss2005}
Mestel L, Moss D {\it Mon. Not. Roy. Astron. Soc.} {\bf 361} 595 (2005)
\bibitem{Istomin} Istomin Ya N in {\it Progress in Neutron Star Research},
A.P.Wass (Ed.), (Nova Science Publisher, New-York, 2005)
\bibitem{Barsukov2010}
Barsukov D P, Tsygan A I {\it Mon. Not. Roy. Astron. Soc.} {\bf 409} 1077 (2010)
\bibitem{BB}
Biryukov A, Beskin G, Karpov S {\it Mon. Not. Roy. Astron. Soc.} {\bf 420} 103 (2012)
\bibitem{Book}
Beskin V S {\it MHD Flows in Compact Astrophysical Objects}
(Springer-Verlag, Berlin, 2010)
\bibitem{GJ}
Goldreich P, Julian W H  {\it ApJ}  {\bf 157} 869 (1969)
\bibitem{LL2}
Landau L D, Lifshits E M, Pitaevskii L P {\it Electrodynamics of Continuous Media} (Oxford: Pergamon Press, 1984)
\bibitem{Schwinger1998}
Schwinger J, Deraad L L, Milton K A, Tsai W, Norton J {\it Classical Electrodynamics}
(Westview Press, 1998)
\bibitem{TPL}
Philippov A, Tchekhovskoy A,  Li J  {\it Mon. Not. Roy. Astron. Soc.} {\bf 441} 1879 (2014)
\bibitem{anapole}
Dubovik V M, Tugushev V V {\it Phys. Rep.} {\bf 187} 145 (1990)
\bibitem{sokolov}
Sokolov I V {\it Sov. Phys. Usp.} {\bf 34} 925 (1991)
\bibitem{chrip}
Bliokh K Yu, Bliokh Yu P {\it Phys. Rev. Lett.} {\bf 96} 073903 (2006)
\end{thebibliography}
\end{document}